\newcommand{\HII}{H\,{\scriptsize II}}

\newcommand{\CII}{[C\,{\scriptsize II}]}

\newcommand{\OI}{[O\,{\scriptsize I}]}

\documentclass[letter]{aa}
\usepackage{graphicx}
\usepackage{txfonts}
\usepackage[draft]{hyperref}
\begin{document} 


   \title{The SOFIA FEEDBACK [CII] Legacy Survey: Rapid molecular cloud dispersal in RCW~79 }


   \author{L. Bonne \inst{1}
   \and S. Kabanovic\inst{2}
   \and N. Schneider\inst{2}
   \and A. Zavagno \inst{3,4} 
   \and E. Keilmann \inst{2} 
   \and R. Simon \inst{2}
   \and C. Buchbender \inst{2} 
   \and R. G\"usten \inst{5} 
   \and A.M. Jacob \inst{5,6}
   \and K. Jacobs \inst{2} 
   \and U. Kavak \inst {1}
   \and F.L. Polles\inst{1} 
   \and M. Tiwari \inst{5} 
   \and F. Wyrowski \inst{5} 
   \and A.G.G.M. Tielens \inst{7,8} 
    }

   \institute{SOFIA Science Center, USRA, NASA Ames Research Center, Moffett Field, CA 94 045, USA 
   \email{lbonne@usra.edu}
   \and{I. Physik. Institut, University of Cologne, Z\"ulpicher Str. 77, 50937 Cologne, Germany} 
   \and{Aix Marseille Univ, CNRS, CNES, LAM, Marseille, France}  
   \and{Institut Universitaire de France, 1 rue Descartes, Paris, France} 
   \and{Max-Planck Institut f\"ur Radioastronomie, Auf dem H\"ugel 69, 53121 Bonn, Germany} 
   \and{Department of Physics \& Astronomy, Johns Hopkins University, Baltimore, MD 21218, USA} 
   \and{University of Maryland, Department of Astronomy, College Park, MD 20742-2421, USA} 
   \and{Leiden Observatory, PO Box 9513, 2300 RA Leiden, The Netherlands} 
     }

\date{draft of \today}

\titlerunning{Cloud dispersal in RCW79}  
\authorrunning{L. Bonne}  

 
\abstract
{It has long been discussed whether stellar feedback in the form of winds and/or radiation can shred the nascent molecular cloud, thereby controlling the star formation rate. However, directly probing and quantifying the impact of stellar feedback on the neutral gas of the nascent clouds is challenging. We present an investigation of this impact toward the RCW~79 \HII\ region using the ionized carbon line at 158 $\mu$m (\CII) from the FEEDBACK Legacy Survey.  
We combine this data with information on the dozen ionizing O stars responsible for the evolution of the region, and observe in \CII\ for the first time both blue- and redshifted  high-velocity gas that reaches velocities of up to 25 km s$^{-1}$ relative to the bulk emission of the molecular cloud. This high-velocity gas mostly contains neutral gas, and partly forms a fragmented shell, similar to recently found shells in a few Galactic \HII\ regions. However, this shell does not account for all of the observed neutral high-velocity gas. We also find high-velocity gas streaming out of the nascent cloud through holes, and obtain a range of dynamical timescales below 1.0 Myr for the high-velocity gas that is well below the 2.3$\pm$0.5 Myr age of the OB cluster. This suggests a different scenario for the evolution of RCW~79, where the high-velocity gas does not solely stem from a spherical expanding bubble, but also from gas recently ablated at the edge of the turbulent molecular cloud into the surrounding interstellar medium through low-pressure holes or chimneys.
The resulting mass ejection rate estimate for the cloud is 0.9-3.5$\times$10$^{-2}$ M$_{\odot}$~yr$^{-1}$, which leads to short erosion timescales ($<$5 Myr) for the nascent molecular cloud. This finding provides direct observational evidence of rapid molecular cloud dispersal. }

\maketitle
%

\section{Introduction} \label{sec:intro}
Feedback from OB stars, in the form of radiation and stellar winds, injects copious amounts of thermal and mechanical energy in the interstellar medium (ISM). As a result, feedback plays a central role in the evolution of the ISM, and might even maintain the low star formation rate observed in molecular clouds \citep{Gao2004,Evans2009,Lada2010}. 

\begin{figure*}[htp]
    \centering
    \includegraphics[width=0.47\hsize]{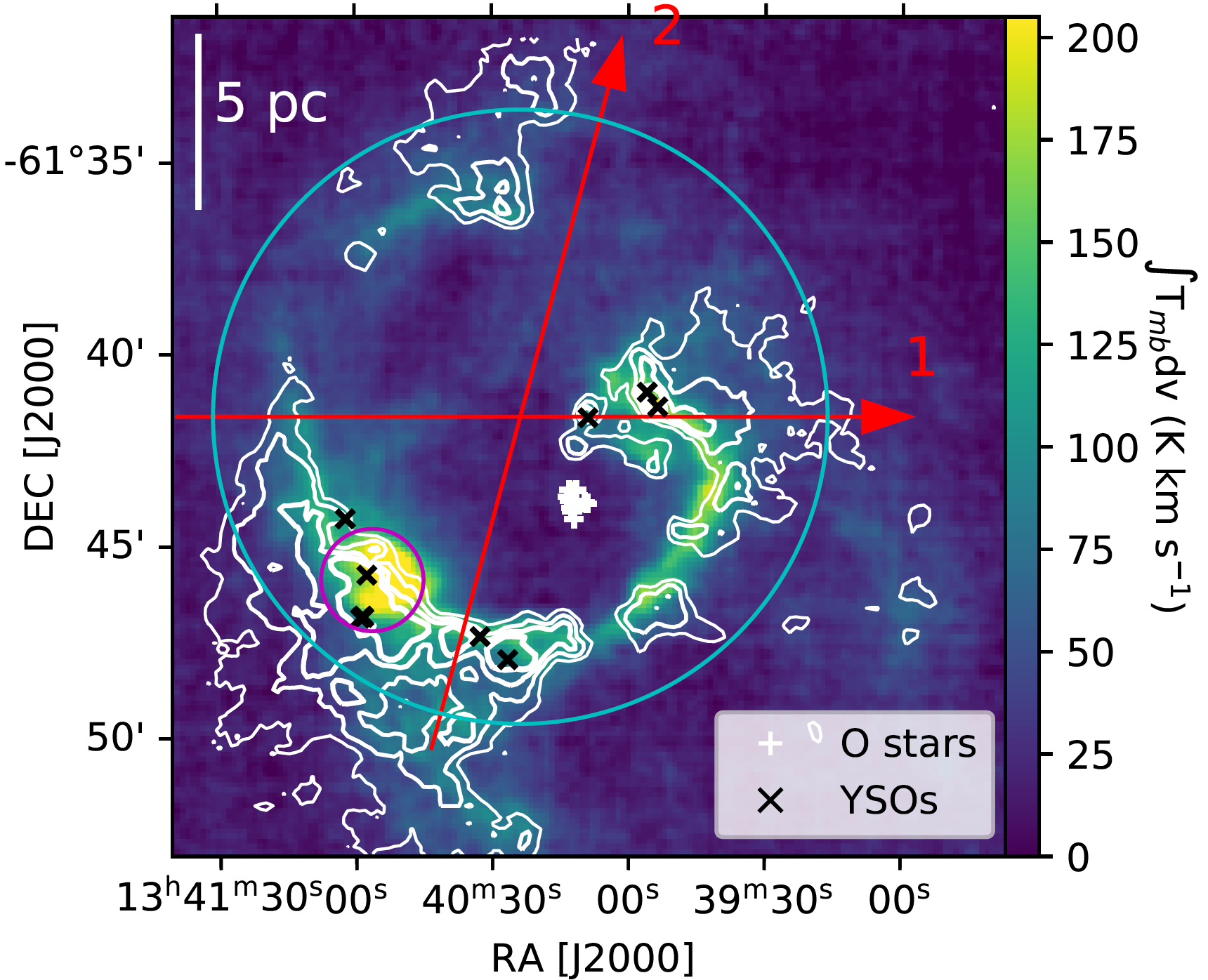}
    \includegraphics[width=0.47\hsize]{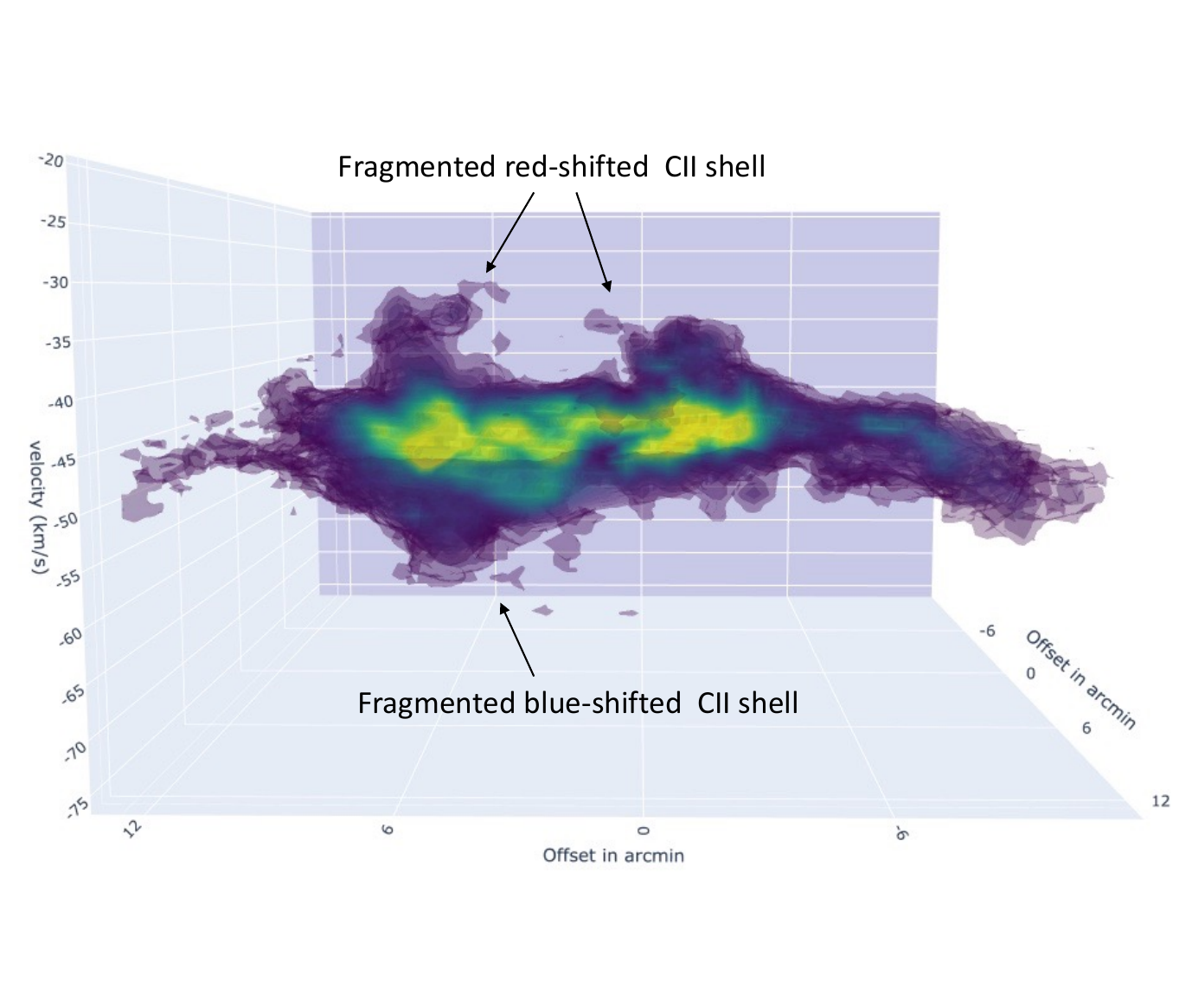}
     \includegraphics[width=0.47\hsize]{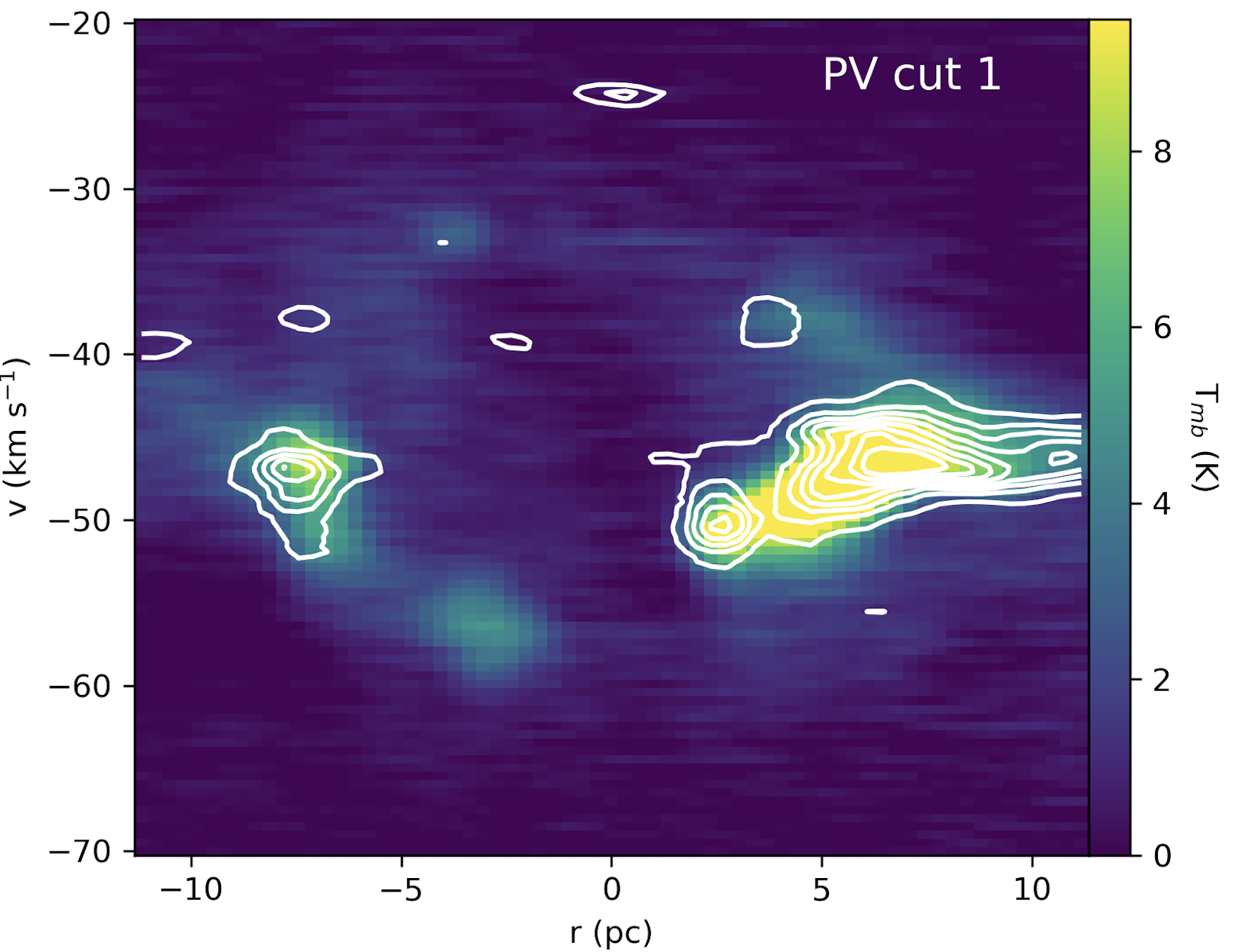}
    \includegraphics[width=0.47\hsize]{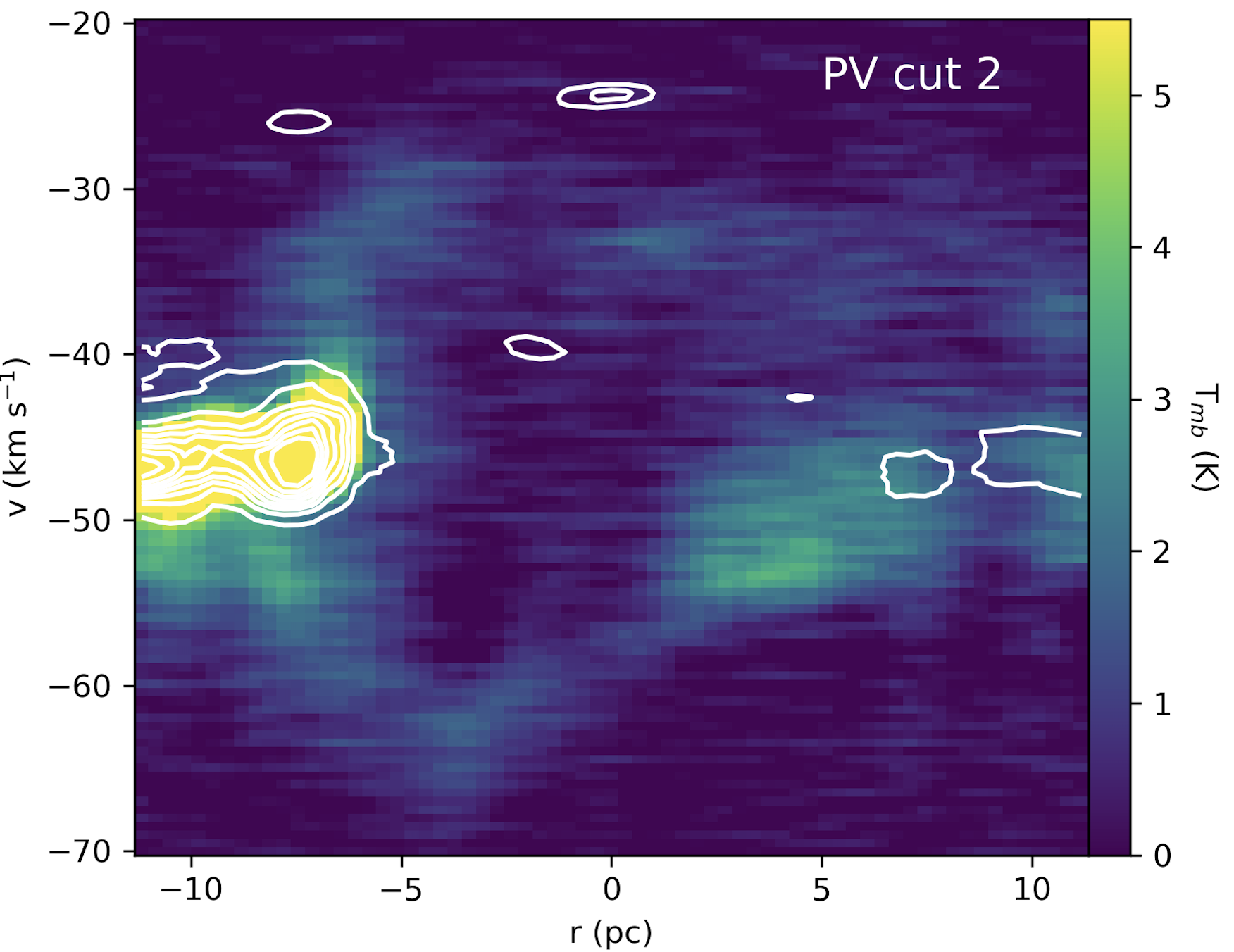}
     \caption{Integrated intensity map and kinematic structure of RCW 79 in \CII. {\bf Top left}:  \CII\ integrated intensity map of RCW~79 between -70 and -20 km s$^{-1}$ with the integrated $^{12}$CO(3-2) emission at 25, 50, 75, and 100 K~km~s$^{-1}$ overlaid by white contours. The white plus signs and black crosses respectively give the locations of the previously identified O stars and YSOs in the region. The red arrows outline the cuts used for the PV diagrams below, with the arrow indicating a positive radius. The cyan circle delimits the region used for the average spectrum presented in Fig. \ref{fig:fig2} and the magenta circle indicates the compact \HII\ region.  Movies of PV cuts over the map are found at \url{https://hera.ph1.uni-koeln.de/~nschneid/rcw79_animations.html}. {\bf Top right}: 2D projection of an interactive 3D isocontour plot, which is found at the same url as above. {\bf Bottom left and right}: \CII\ PV diagrams with the $^{12}$CO(3-2) emission indicated by white contours. Both blue- and redshifted high-velocity \CII\ emission is detected. The high-velocity \CII\ gas is typically found toward the edges of the \HII\ region, and flowing out through the opening in the northwest (see PV cut 2).}
    \label{fig:fig1}
\end{figure*}

The \CII\ fine-structure line at 158 $\mu$m is ideally suited to study the impact of stellar feedback on the embedding molecular cloud as it mostly probes photodissociation regions (PDRs) where far-ultraviolet (FUV) photons between 6 and 13.6 eV dominate the physical and chemical state of the gas \citep{Hollenbach1999,Wolfire2022}. This has made the \CII\ spectral line   a prominent target for observing programs on far-infrared observatories  \citep[e.g.,][]{Stacey1993,Pineda2013,Pineda2014,Goicoechea2015}. Recently, the Stratospheric Observatory for Infrared Astronomy (SOFIA) facilitated efficient velocity-resolved \CII\ mapping around \HII\ regions. This particularly revealed the presence of expanding \CII\ shell morphologies at high velocity (up to $\sim$15~km~s$^{-1}$). However, these shells were typically found to expand in one direction, mostly  toward the observer  \citep{Pabst2019,Pabst2020,Luisi2021,Tiwari2021,Kabanovic2022,Beuther2022,Bonne2022,Tram2022}. 
Early work by \citet{Pabst2019} and \citet{Luisi2021} attributed these observed dynamic features in \CII\ to a single roughly coherent expanding shell or bubble that is typically considered in analytical models \citep[e.g.,][]{Spitzer1978,Weaver1977}. However, observations in \citet{Beuther2022} and \citet{Bonne2022} showed that \CII\ emission at high velocities can also have a more complex morphology.
Constraining the nature of high-velocity \CII\ gas is thus important in order to understand the impact of stellar feedback on molecular cloud evolution, which will have implications for molecular cloud lifetimes and the long-standing debate of  whether clouds evolve dynamically or in a quasi-static fashion \citep[e.g.,][] {Shu1987,Elmegreen2000,Hartmann2001,Krumholz2005,Schneider2023}.\\ 
Here we present a currently unique combination of \CII\ observations from the SOFIA legacy survey FEEDBACK  \citep{Schneider2020} and the results from a detailed study of the ionizing O stars \citep{Martins2010} for the \HII\ region RCW~79 (see   Fig. \ref{fig:fig1}). RCW~79 has been located at distances between 4.0 and 4.3 kpc \citep{Russeil1998,Mege2021};  here we update the distance to 3.9$\pm$0.4 kpc using the GAIA DR3 release (see Appendix \ref{sec:GAIA}). The region is heated by a cluster of 12 identified O stars, including O4-6V/III candidates, that have an estimated age of 2.3$\pm$0.5 Myr \citep{Martins2010}. Gas compression by stellar feedback has created a fragmented quasi-circular ring of dense molecular gas \citep{Zavagno2006,Liu2017} with a radius of 6-7 pc. This ring hosts 
young stellar objects (YSOs) and one compact \HII\ region in the southeast that were proposed by \citet{Zavagno2006} to be the result of the collect and collapse mechanism \citep{Elmegreen1977}. Figure \ref{fig:fig1} shows that the O stars are located in the southwestern area of the \HII\ region. 

\begin{figure}
    \centering
    \includegraphics[width=\hsize]{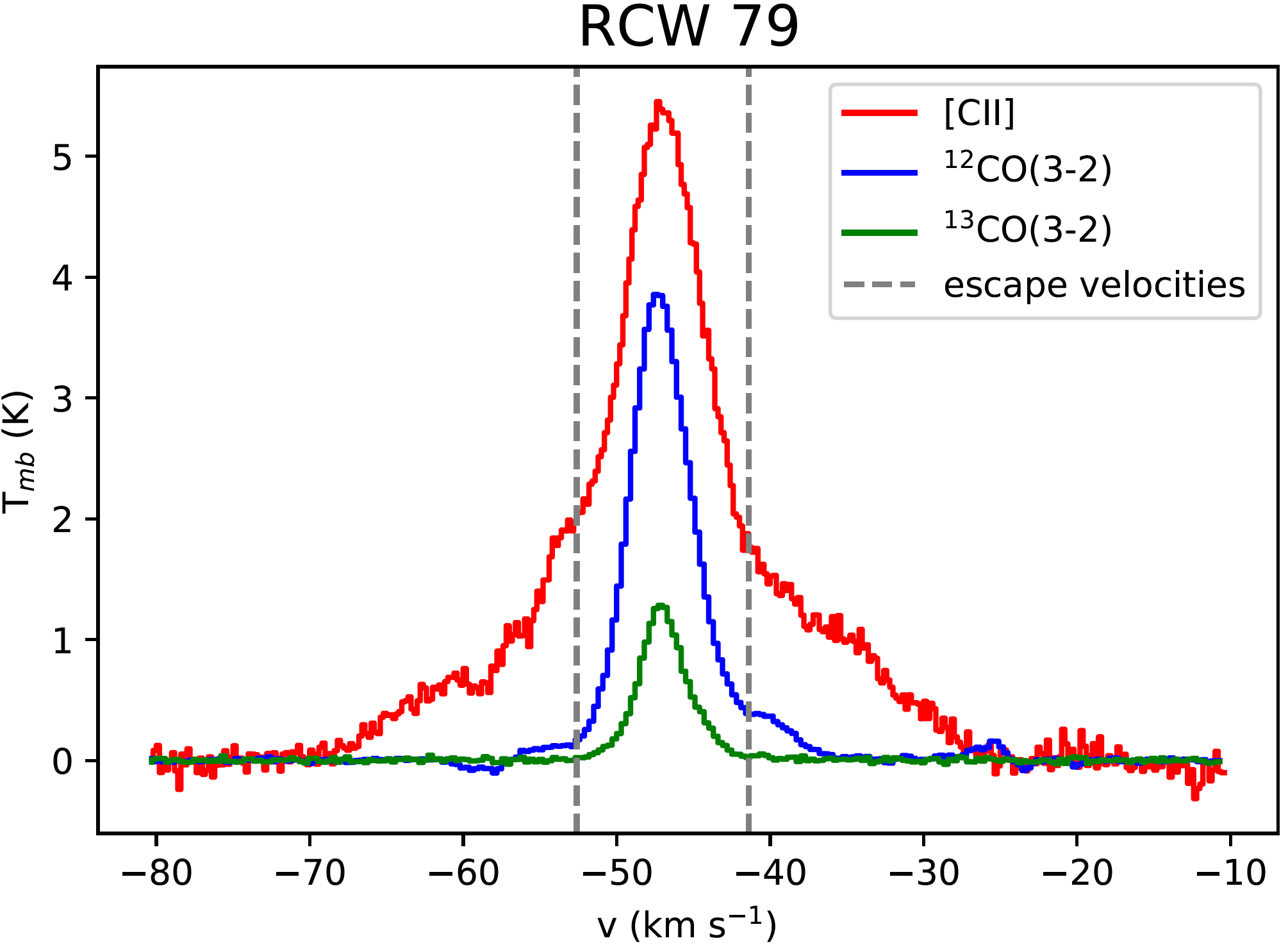}
    \caption{Average \CII, $^{12}$CO(3-2), and $^{13}$CO(3-2) spectrum toward RCW 79. 
    The gray dashed vertical lines indicate the escape velocities of the molecular cloud relative to its central velocity of -47~km~s$^{-1}$. The \CII\ spectrum has evident high-velocity wings outside the escape velocity range of the molecular cloud.
    }
    \label{fig:fig2}
\end{figure}

\section{Observations}

\subsection{SOFIA}
The \CII\ $^{3}$P$_{3/2} \rightarrow$ $^{3}$P$_{1/2}$ line at 158 $\mu$m was observed in parallel with the \OI\ 63 $\mu$m line with the upgraded German REceiver for Astronomy at Terahertz frequencies (upGREAT) \citep{Risacher2018} on board SOFIA \citep{Young2012}. An area of $\sim$470 arcmin$^2$ was mapped in the on-the-fly (OTF) mode and atmospheric calibration was done with the GREAT pipeline \citep{Guan2012}. Including calibration, this required a total observing time of 10 h. The antenna temperature was converted to main beam temperature using a forward efficiency $\eta_{for}$ = 0.97 and a main beam efficiency $\eta_{mb}$ = 0.65. To improve the data quality and remove scanning effects in the form of stripes, we employed a method based on principal component analysis (PCA) that was also used for the data presented in other FEEDBACK papers \citep{Tiwari2021,Kabanovic2022,Schneider2023,Bonne2023b}. The nominal angular resolution of the \CII\ and \OI\ data is 14.1$''$ and 6$''$, respectively, but here we convolve the \CII\ data cube to an angular resolution of 20$^{\prime\prime}$ and a spectral binning of 0.5~km~s$^{-1}$. This slight smoothing and rebinning improves the quality of the data cube, for example by reducing the noise rms and striping effects. The noise rms in one channel is typically 0.5--0.6 K (see \citealt{Schneider2020} for more observational details, such as the OFF position). 


\subsection{APEX}
The $^{12}$CO(3-2) and $^{13}$CO(3-2) transitions at 345.796~GHz and 330.588~GHz, respectively, were mapped over the same region as the \CII\ map with the LAsMA receiver on the APEX telescope \citep{Guesten2006}. The observations were done in the total power OTF mode, and the antenna temperature was converted to main beam temperature using $\eta_{mb}$ = 0.68. The final data cubes were produced with first-order baseline removal with an angular resolution of $\sim$20$''$ and a spectral binning of 0.3 km s$^{-1}$. This results in a typical noise rms of 0.4-0.5 K in one channel. 

\begin{figure}
    \centering
    \includegraphics[width=0.85\hsize]{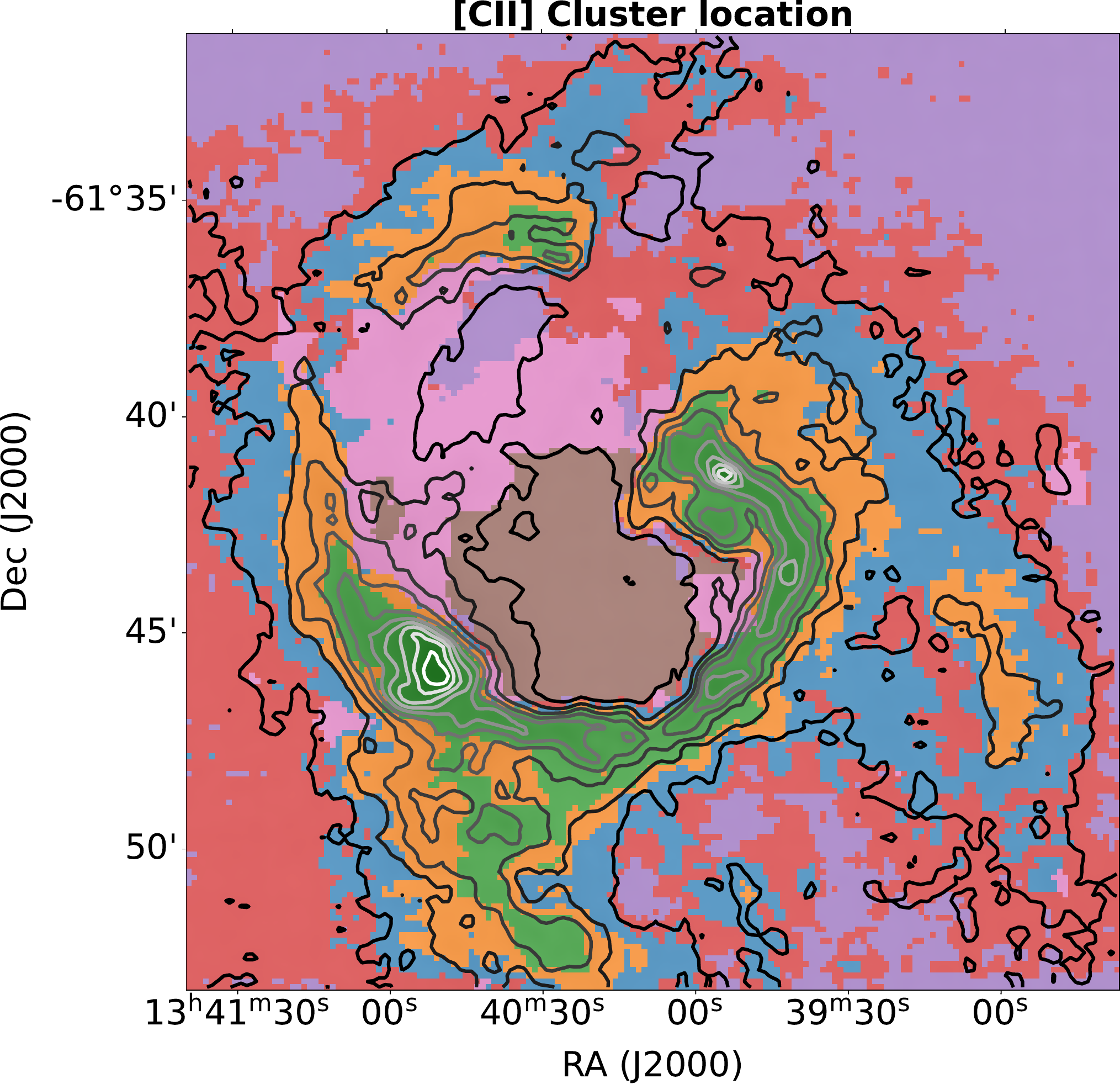}
    \includegraphics[width=0.85\hsize]{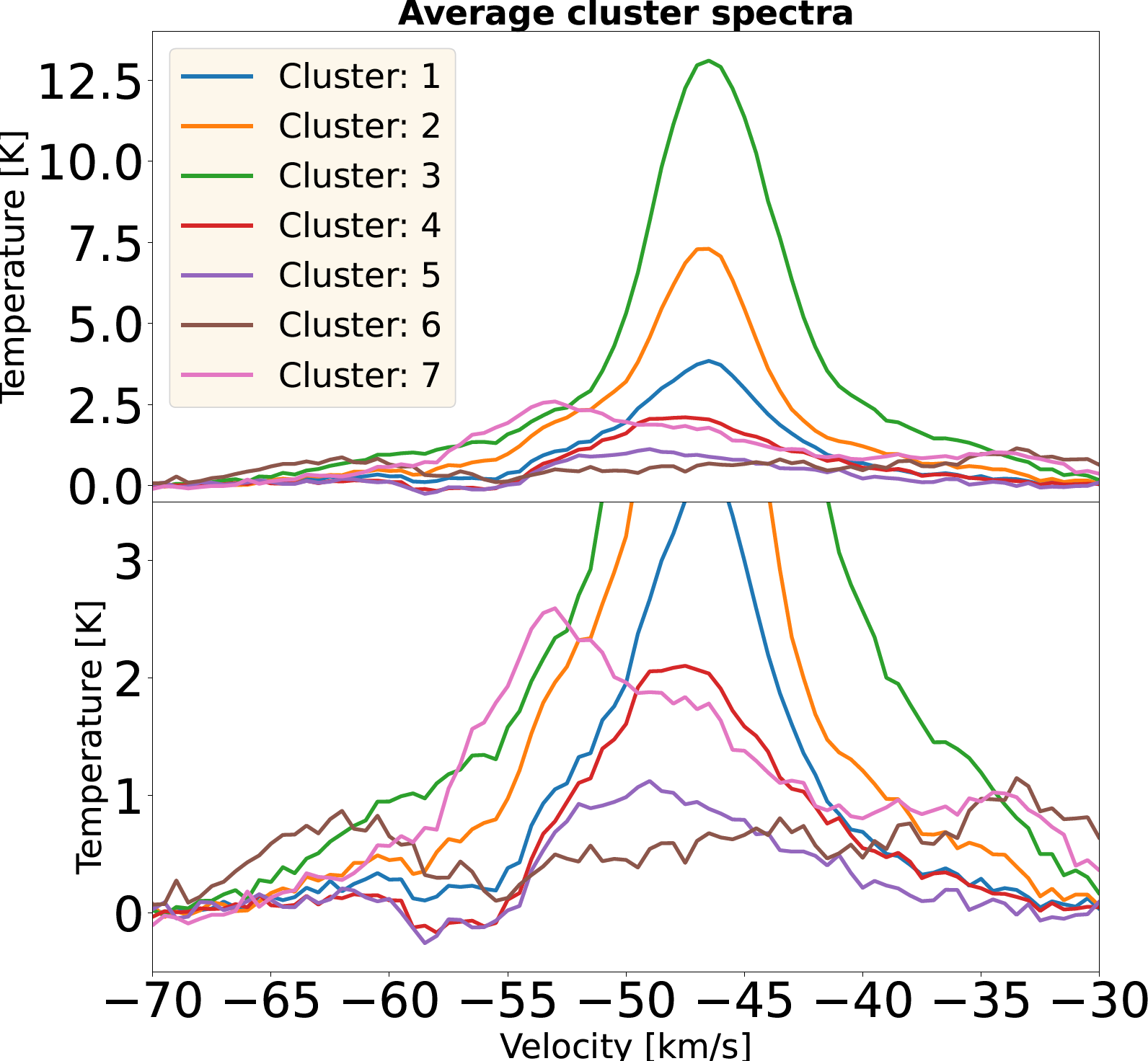}
    \caption{Output of the unsupervised  GMM clustering algorithm on the \CII\ data of RCW~79. {\bf Top}: Cluster locations in RCW 79, representing zones with similar \CII\ spectral line profiles, identified with the GMM. The contours indicate the total \CII\ integrated intensity, starting at 25 K~km~s$^{-1}$ with increments of 25 K~km~s$^{-1}$, revealing that the regions identified based on the normalized \CII\ spectral morphology with the GMM are clearly correlated to spatial features identified in the intensity map of the region. {\bf Bottom}: Corresponding average spectra for each cluster identified by the GMM calculations on RCW~79.}
    \label{fig:fig2a}
\end{figure}

\section{Results}

\subsection{Integrated intensity map and spectra}
\begin{figure}
    \centering
    \includegraphics[width=\hsize]{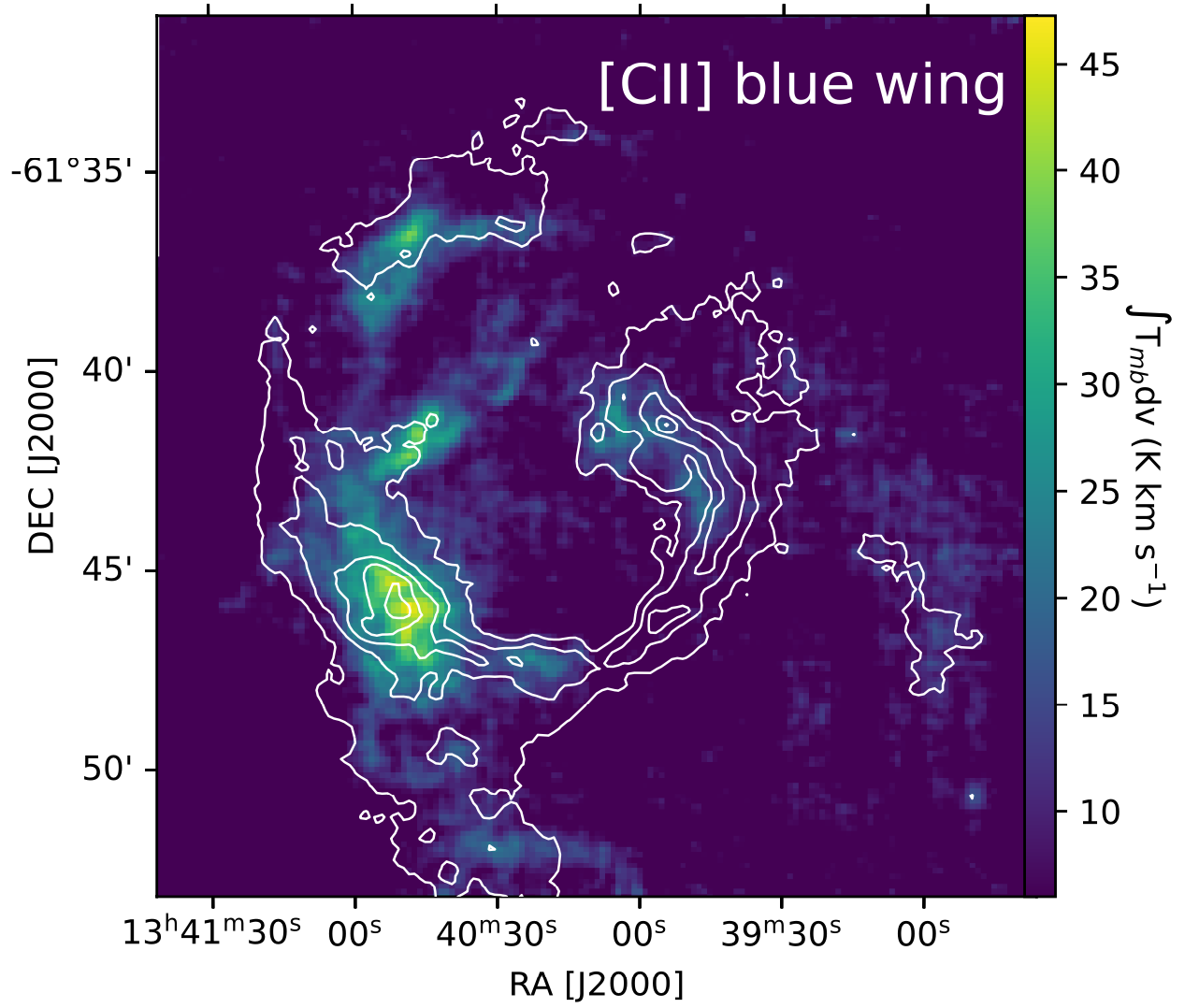}
    \includegraphics[width=\hsize]{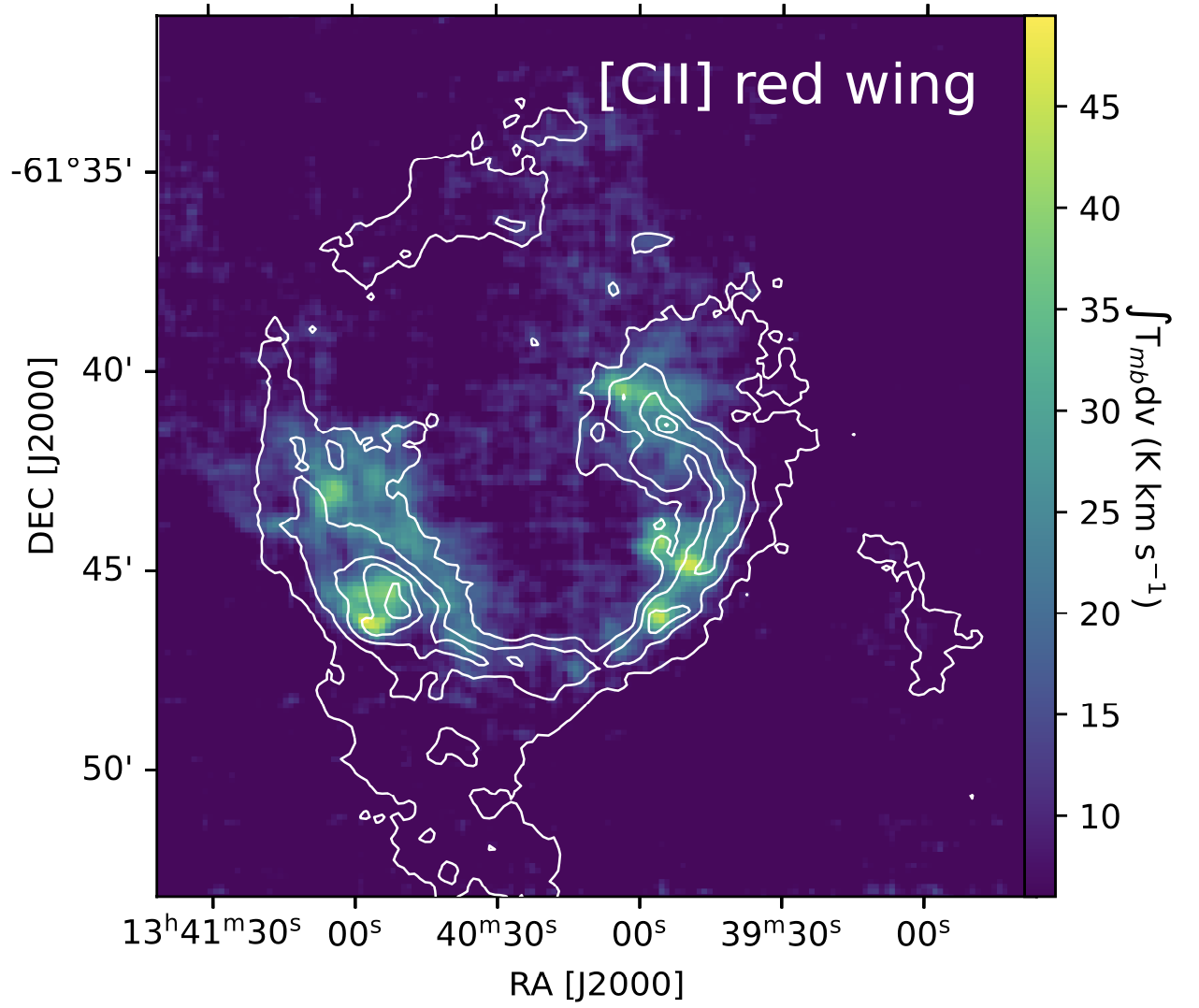}
    \caption{Spatial distribution of the blue- and redshifted \CII\ high-velocity gas in RCW~79. {\bf Top}: Intensity map, integrated over the blueshifted high-velocity wing (i.e., $-$70 to -52.6 km s$^{-1}$). The high-velocity gas is also observed to the south of RCW~79. The white contours indicate the total integrated \CII\ emission (presented in Fig. \ref{fig:fig1}) starting at 50~K~km~s$^{-1}$ with increments of 50~K~km~s$^{-1}$. {\bf Bottom}: Same, but for the redshifted high-velocity wing (i.e., $-$41.4 to -25 km s$^{-1}$). The high-velocity gas is observed passing through the openings of the \HII\ region in the northeast and northwest.}
    \label{fig:outflowMaps}
\end{figure}

The \CII\ integrated intensity map of RCW~79 in Fig.~\ref{fig:fig1} (top left) is dominated by a  circular morphology very similar to that observed in the Spitzer infrared maps presented in \citet{Zavagno2006}. Most of the \CII\ emission is located at the interface of the \HII\ region and the $^{12}$CO emission from the nascent molecular cloud, and has a brightness peak at the location of a compact \HII\ region in the southeast. However, there is also emission outside of the ring, which is discussed below, and low-surface brightness emission inside the \CII\ ring. Figure~\ref{fig:fig2} presents the spatially averaged spectrum over the \HII\ region which reveal  prominent high-velocity wings in \CII\ between $-$70 and -52.6 km s$^{-1}$ (blueshifted)   and $-$41.4 and -25 km s$^{-1}$ (redshifted), that are mostly undetected in $^{12}$CO. These velocities are outside the escape velocity\footnote{v$_{esc,0}$ = $\sqrt{2\,G\,M_{cloud,0}/r_{cloud,0}}$ with G the gravitational constant, M$_{cloud,0}$ (= 5.5$\times$10$^{4}$ M$_{\odot}$) the original mass of the cloud (i.e., the current mass of the cloud corrected for the mass ejection that has been going on for 2.3$\pm$0.5 Myr; see Sect. \ref{sec:massEjSec}), and r$_{cloud,0}$ (= 15 pc) the estimated original size of the cloud.} range of the molecular cloud. Such \CII\ high-velocity wings were detected early on in spectrally resolved \CII\ observations of \HII\ regions \citep[][]{Simon2012,Schneider2018}, but their nature has not been discussed in detail.\\
The different integrated emission features are also clearly identified when applying the Gaussian mixture model (GMM) methodology, introduced in \citet{Kabanovic2022}, to the \CII\ data cube (see Fig. \ref{fig:fig2a}). In short, GMM sorts spectra into clusters with similar line profiles by describing the spectral parameters\footnote{The spectral parameters are the intensity of each spectral bin in the spectra.} of all spectra in the data cube with a linear combination of Gaussian distributions (see Sect. 4.3.1 in \citealt{Kabanovic2022} for a detailed description of the method). The brightness range of all spectra is normalized so that GMM clusters the spectra based on the line profile regardless of the absolute intensity.  
We then employ the Bayesian information criterion (BIC) to determine the best fitting number of clusters. Figure~\ref{fig:fig2a} displays the spatial distribution of the seven identified clusters in RCW~79, and  shows the spatial association of kinematic substructure in the region. The bright \CII\ ring is separated into three clusters: cluster 1 (blue), 2 (orange), and 3 (green) where the spectra from clusters 2 and 3, which trace the inner regions of the ring, have high-velocity wings. The wings in cluster 2 are predominantly blueshifted, while cluster 3 has equally bright blue- and redshifted wings. Clusters 4 (red) and 5 (purple) are located outside of the PDR ring and have slightly more blueshifted velocities. Clusters 6 (brown) and 7 (pink) probe the region inside the ring with brightness peaks at more extreme blue- and redshifted velocities (up to $-$64 km s$^{-1}$ and $-$34 km s$^{-1}$, respectively).\\
In Fig. \ref{fig:outflowMaps} we present the line emission integrated over the blue- and redshifted high-velocity wings. This emission is unevenly distributed for the blue- and redshifted gas along the ring that identifies the RCW~79 bubble, indicating that their structure is affected by the 3D morphology of the nascent cloud. We also note that  the blue- and the redshifted high-velocity gas is observed outside of the \HII\ region. The blueshifted gas is mostly observed south of the region, while the redshifted gas is breaking through the northwestern and northeastern holes in the ring.

\subsection{Position-velocity diagrams}

The position-velocity (PV) diagrams and movies in Fig.~\ref{fig:fig1} present the \CII\ and $^{12}$CO(3-2) kinematics over the \HII\ region. The PV cut 1 in Fig.~\ref{fig:fig1} shows the high-velocity gas structure in \CII\ that is expected for a spherical expanding shell. It is also the first time that we simultaneously clearly observe both a blue- and redshifted shell morphology. Cut 2, on the other hand, was chosen as the axis goes through a region in the northwest where the ring has clearly broken open. This shows that the \CII\ high-velocity gas is not spherically expanding and that high-velocity gas is flowing out of the region through the opening. 
The interactive 3D isocontour plot in Fig.~\ref{fig:fig1} and the PV movies show the \CII\ velocity structure in more detail, which further demonstrates the complex structure of the high-velocity gas. This complexity is also demonstrated by the GMM analysis that identifies four distinct clusters with high-velocity gas (see previous section). In addition, we also find that the spatial distribution of the high-velocity \CII\ emission is inconsistent with a single spherical expanding bubble (see Appendix \ref{sec:limb}).
 
\section{Discussion}

\begin{figure}
    \centering
    \includegraphics[width=\hsize]{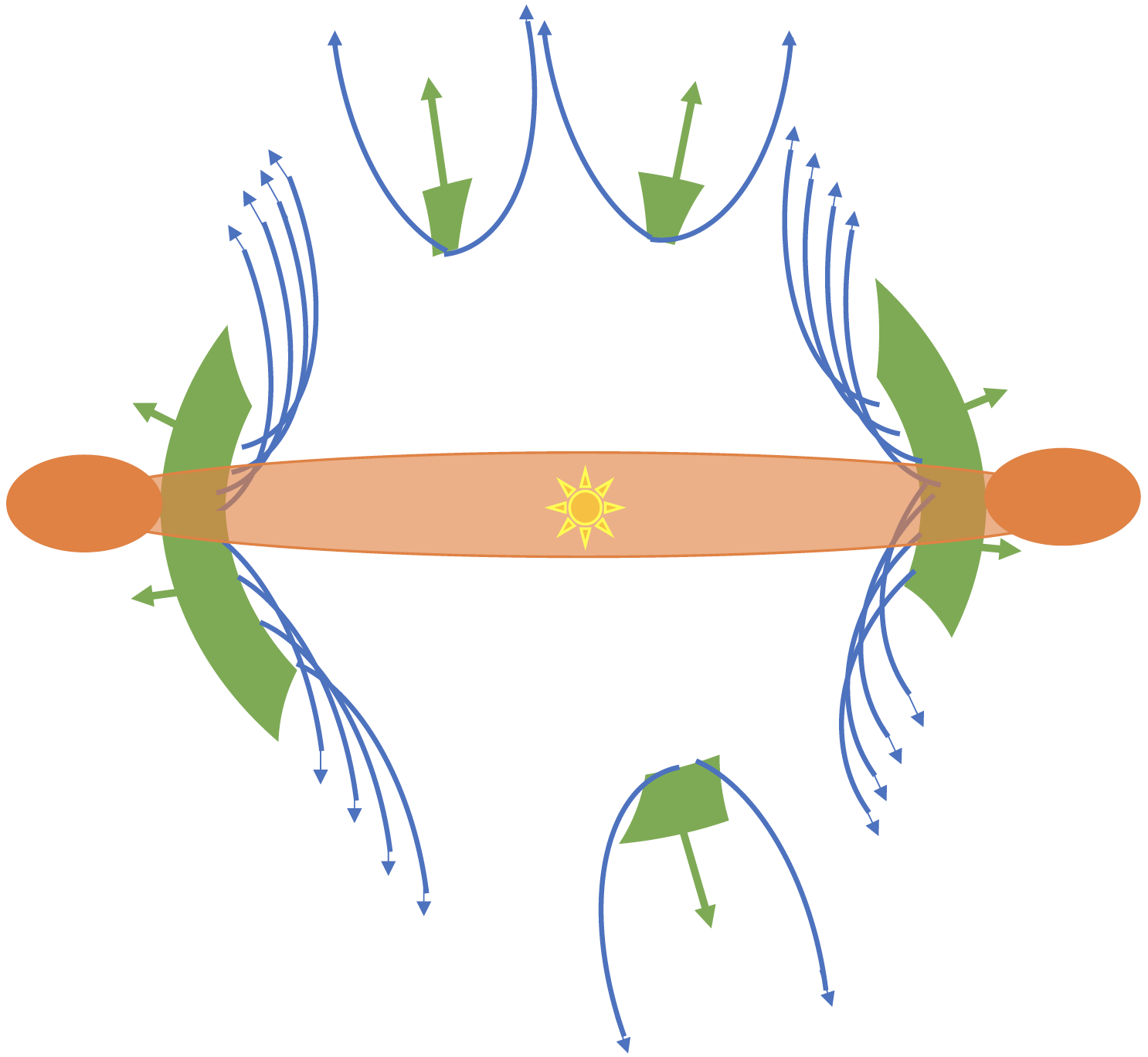}
    \caption{Schematic illustration of the proposed scenario for RCW~79, seen edge-on (i.e., from in the plane of the sky). The green ring indicates the expanding fragmented shells with multiple holes. Through these holes, high-velocity neutral gas, indicated with the blue arrows, is escaping into the ambient ISM. Also shown  (in orange) is the ring-like structure of molecular gas, wrapping the \HII\ region,   seen in CO from an observing position in the plane of the sky.}
    \label{fig:sketchHoles}
\end{figure}

\subsection{High-velocity mass ablation in RCW 79}
To further discuss the observed  \CII\ high-velocity gas, we need to constrain the 3D distance of the \CII\ high-velocity emission region from the ionizing O star cluster. For this we use the integrated \CII\ brightness in the wings, which has values spread over the observed region from the 3$\sigma$ detection limit ($\sim$8 K km s$^{-1}$) up to 50 K km s$^{-1}$ for each wing. Using the PDR Toolbox \citep{Pound2022}, this points to a typical strength of G$_{0}$ $\approx$ 100-1000 for the FUV field that excites the \CII\ emission in the wings (assuming a density n$_{H}\sim$10$^{2}$-10$^{4}$ cm$^{-3}$). 

The stellar parameters from \citet{Martins2010} allow us to calculate the FUV field strength as a function of distance from the cluster.\footnote{This method follows the r$^{-2}$ decrease in FUV intensity, integrated between 6 and 13.6 eV without dust attenuation, from the surface of the stellar objects (see \citealt{Bonne2020} for more details).} This results in 3D distances from the O-star cluster of 5-12 pc for the \CII\ emission in the wings.\footnote{This fits with the observation of high-velocity wings in RCW 36. These observations show that the 3D extent of the high-velocity wings was within a factor two of the radius of the swept-up ring \citep{Bonne2022}.} Based on the observed plane-of-the-sky (POS) radius this gives a distance d in the line of sight (LOS) between 0 and 10 pc, which allows us to estimate the dynamical timescales (t$_{dyn}$ = d$_{LOS}$/v$_{wing}$). 
Using a typical d$_{LOS}$ = 5 pc and the velocity range of the \CII\ wings relative to the cloud for v$_{wing}$ then results in a full range of dynamical timescales of $\approx$0.2-1.0 Myr. This is significantly shorter than the observationally calculated cluster age of 2.3$\pm$0.5 Myr \citep{Martins2010} and an age estimate of 2.2$\pm$0.1 Myr by \citet{Tremblin2014}. This suggests that the high-velocity gas is continuously replenished as it flows away, and that \CII\ thus only traces the gas that is currently being ejected from the region ($<$ 1 Myr old). Combined with the observed complex morphology of the high-velocity gas, we argue that the high-velocity gas is not a single expanding shell in RCW~79, in contrast to what has been observed for several other \HII\ regions \citep{Luisi2021,Pabst2019}. We propose that the high-velocity \CII\ emission also originates from gas that is continuously entrained at the interface of the \HII\ region and the dense ring and turbulent molecular cloud. This entrained gas escapes from the molecular clouds through low-density holes in the turbulent cloud (see Fig. \ref{fig:sketchHoles}). In this context it is interesting to note that observations in the Orion cloud found that protostellar outflows can also create holes in an expanding shell \citep{Kavak2022a,Kavak2022b}. In addition to the turbulent structure of the cloud, protostellar outflows could thus also contribute at the early feedback stages in the creation of these holes. However, there are currently no clear indications of an important role for protostellar outflows in RCW~79.
We also note that \citet{Zavagno2006} suggested that ionized gas from the \HII\ region is flowing outward based on their H$\alpha$ observations. The scenario for RCW~79 is conceptually similar to that of mass erosion from a ring of dense gas that was recently described in \citet{Whitworth2022}.  

\subsection{Mass ejection rate and cloud erosion timescale}\label{sec:massEjSec}
With the estimate of the dynamic timescales for the high-velocity wings and the calculated C$^{+}$ column density in the velocity channel of the wings  we derive a mass ejection rate of 0.9-3.5$\times$10$^{-2}$ M$_{\odot}$ yr$^{-1}$ associated with the \CII\ wings (see Appendix \ref{sec:appB} for details). This range of values is in agreement with the mass ejection rate ($\sim$10$^{-2}$ M$_{\odot}$ yr$^{-1}$) found in the simulations by \citet{Walch2012,Walch2013}, with cluster and cloud properties that closely match RCW~79. Interestingly, these simulations also contain a variety of expanding structures that fit   the observed range of dynamical timescales of the observed mass ejection. From the Herschel dust column density map\footnote{The Herschel dust column density map at 18$''$ angular resolution of the RCW~79 region \citep{Liu2017} was produced from the HOBYS survey \citep{Motte2010} using the method described in \citet{Palmeirim2013}.} in \citet{Liu2017}, we know that most of the dense gas is located in the swept-up ring at a radius of 6-7~pc with ongoing star formation. However, we consider a radius of 15~pc to include most of the mass of the molecular cloud, which gives a current total cloud mass of 2.1$\times$10$^{4}$ M$_{\odot}$. Combined with the mass ejection rates, this gives a current cloud erosion timescale of 0.6-2.3 Myr. With the estimated cluster age of 2.3$\pm$0.5 Myr, we thus obtain an erosion timescale for the original molecular cloud that is $<$5 Myr. Focusing only on the swept-up ring, within a radius of 8 pc, gives a current erosion timescale of 0.2-0.9 Myr (which adds up to a total erosion time of 2.5-3.2 Myr with the cluster age). This assumes that the high-velocity mass ejection from the cloud is continuous, which is supported by the presence of high-velocity wings in all the targets of the FEEDBACK survey, and in  other \HII\ regions (Bonne et al. in prep.). It is interesting to note that this direct estimate of the molecular cloud erosion timescale fits with numerical simulations of stellar feedback in turbulent clouds \citep{Kim2018,Kim2021b} and indirect observational estimates of the molecular cloud erosion timescales in the Milky Way and nearby galaxies based on extinction measurements and the molecular gas distribution near OB clusters \citep[e.g.,][]{Leisawitz1988,Leisawitz1989,Hannon2019,Kim2021a,Chevance2022}.

\section{Conclusion}
We   presented high spectral resolution \CII\ observations of the \HII\ region RCW~79 from the SOFIA legacy project FEEDBACK. The \CII\ emission traces the PDRs at the edge of the \HII\ region and a compact \HII\ region in the southeast. Most importantly, the map reveals a large quantity of high-velocity gas. This gas displays an expanding shell morphology over the region along a few axes, where   for the first time we identify both a blue- and a redshifted expanding \CII\ shell. However, the overall kinematic structure of this high-velocity gas varies over the \HII\ region with clear evidence of high-velocity gas escaping from the bubble through holes in the nascent cloud that gives rise to a complex dynamic morphology. These proposed holes can naturally emerge in a turbulent cloud, but it cannot be excluded that protostellar outflows might also play a role at earlier stages in their creation. From the velocities of the high-velocity gas we derive a range of dynamical timescales below 1.0 Myr. As these timescales are significantly shorter than the estimated cluster age of 2.3$\pm$0.5 Myr \citep{Martins2010}, we propose that the \CII\ wings only trace the most recent gas that is ablated from the dense ring and ejected from the turbulent cloud. Quantifying the associated mass ejection rate (0.9-3.5$\times$10$^{-2}$ M$_{\odot}$ yr$^{-1}$) implies relatively short estimated erosion timescales for the initial molecular cloud that are $<$5 Myr, which provides direct observational evidence for rapid molecular cloud evolution in this region, consistent with extragalactic studies. However, the current data does not allow us to draw a conclusion on the physical process that drives the molecular cloud erosion (i.e., whether stellar wind or radiation dominates). Additional observations will be required to address this question.

\begin{acknowledgements}
We thank the anonymous referee for constructive and insightful comments that improved the clarity of this letter. We thank F. Comeron, D. Russeil and F. Martins for fruitful discussions on the determination of the new distance of RCW~79 and how it affects the deduced properties of the ionizing O stars.  We also thank L. Lancaster for interesting discussions, and J-G. Kim for insightful comments on a draft version. This study was based on observations made with the NASA/DLR 
Stratospheric Observatory for Infrared Astronomy (SOFIA). SOFIA is
jointly operated by the Universities Space Research Association
Inc. (USRA), under NASA contract NNA17BF53C, and the Deutsches SOFIA
Institut (DSI), under DLR contract 50 OK 0901 to the University of
Stuttgart. upGREAT is a development by the MPIfR and
the KOSMA/University Cologne, in cooperation with the DLR Institut
f\"ur Optische Sensorsysteme.\\
Financial support for FEEDBACK at the University of Maryland was provided by NASA through award
SOF070077 issued by USRA. The FEEDBACK project is supported by the
BMWI via DLR, project number 50 OR 2217 (FEEDBACK-plus).
S.K. acknowledges support by the BMWI via DLR, project number 50OR2311 (Orion-Legacy). A.Z. acknowledges the support of the IUF.
\end{acknowledgements}

\bibliography{aandaRCW79}

\begin{appendix}
\section{Distance update of RCW~79 with GAIA DR3}\label{sec:GAIA}
With GAIA data release 3 \citep[DR3;][]{Gaia2022}, we updated the distance of the RCW~79 ionizing cluster. From a list of all stars in the region of the RCW 79 cluster with parallax measurements, we selected stars that lie within 1$''$ of the O stars from \citet{Martins2010} and have a distance within 3.2 and 5.2 kpc (which is the largest error bar for the region). This provided us with two candidate stars that correspond to an O star identified in \citet{Martins2010}. These two stars have a consistent distance of 3.87$\pm$0.07 kpc and 3.93$\pm$0.06 kpc, a good astrometric solution, and only a velocity difference of 6~km~s$^{-1}$ that might be associated with internal motion in the cluster. Furthermore, the magnitudes of the stars are consistent with reddened hot massive stars. Based on these values, and taking into account a possible parallax zero-point offset, we propose a new distance of 3.9$\pm$0.4 kpc. As the spectral types in \citet{Martins2010} were derived based on spectral classification, the new distance does not affect these results nor does it affect the upper limit on the stellar wind rates. Only the ionizing photon flux requires a 14\% correction.

\section{Limb brightening in a spherical shell}\label{sec:limb}
\begin{figure}
    \centering
    \includegraphics[width=\hsize]{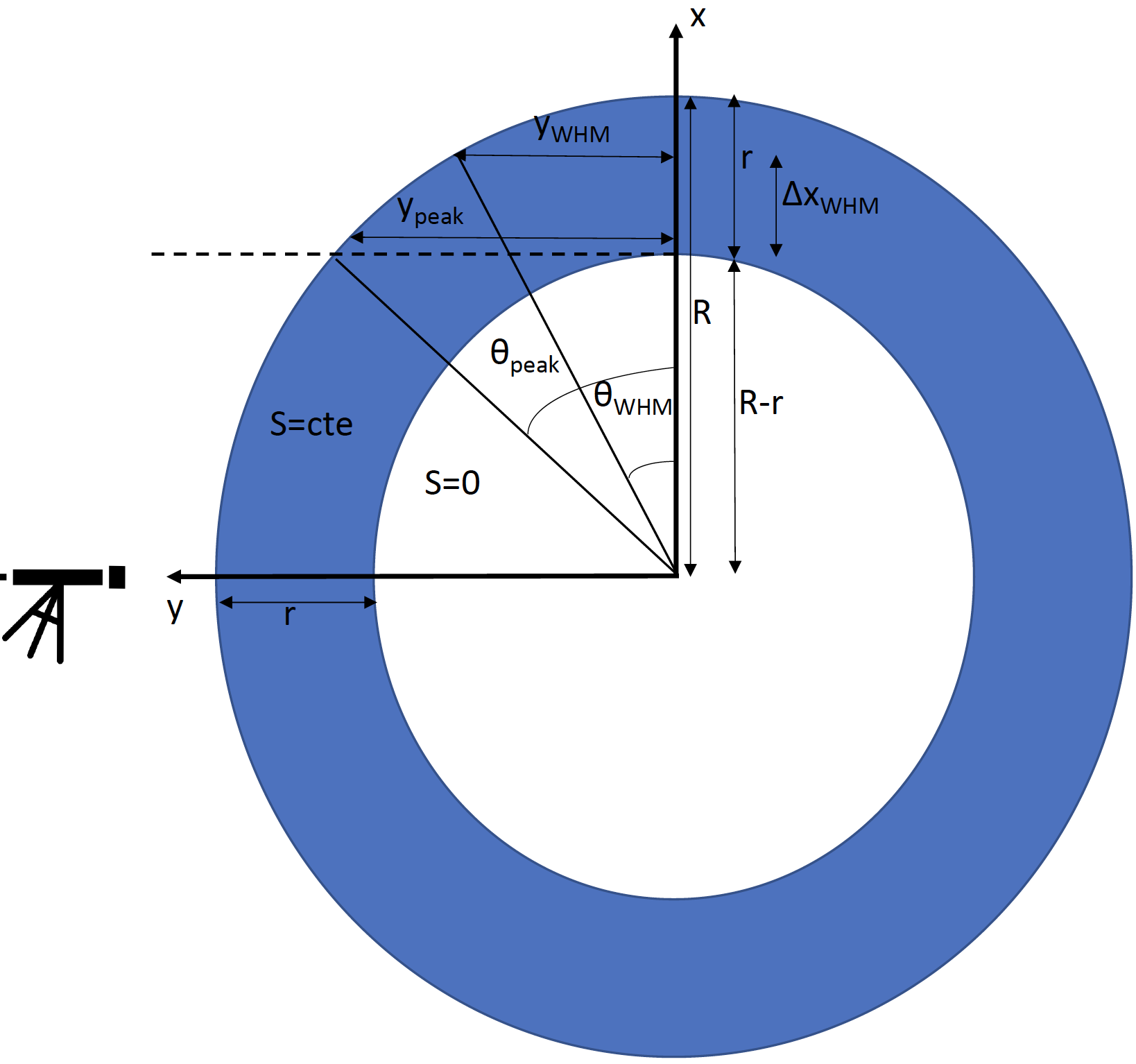}
    \caption{Spherical shell or bubble  observed along the y-axis, as indicated by the telescope illustration. The shell (in blue) surrounding a bubble (white) is shown with a constant specific intensity (S),  radius R, and thickness r. The maximum observed limb brightening is at x = R-r, and is indicated by y$_{peak}$. The observed width at half the intensity in the shell (y$_{WHM}$) is indicated with $\Delta$x$_{WHM}$.}
    \label{fig:limb}
\end{figure}

\begin{figure}
    \centering
    \includegraphics[width=\hsize]{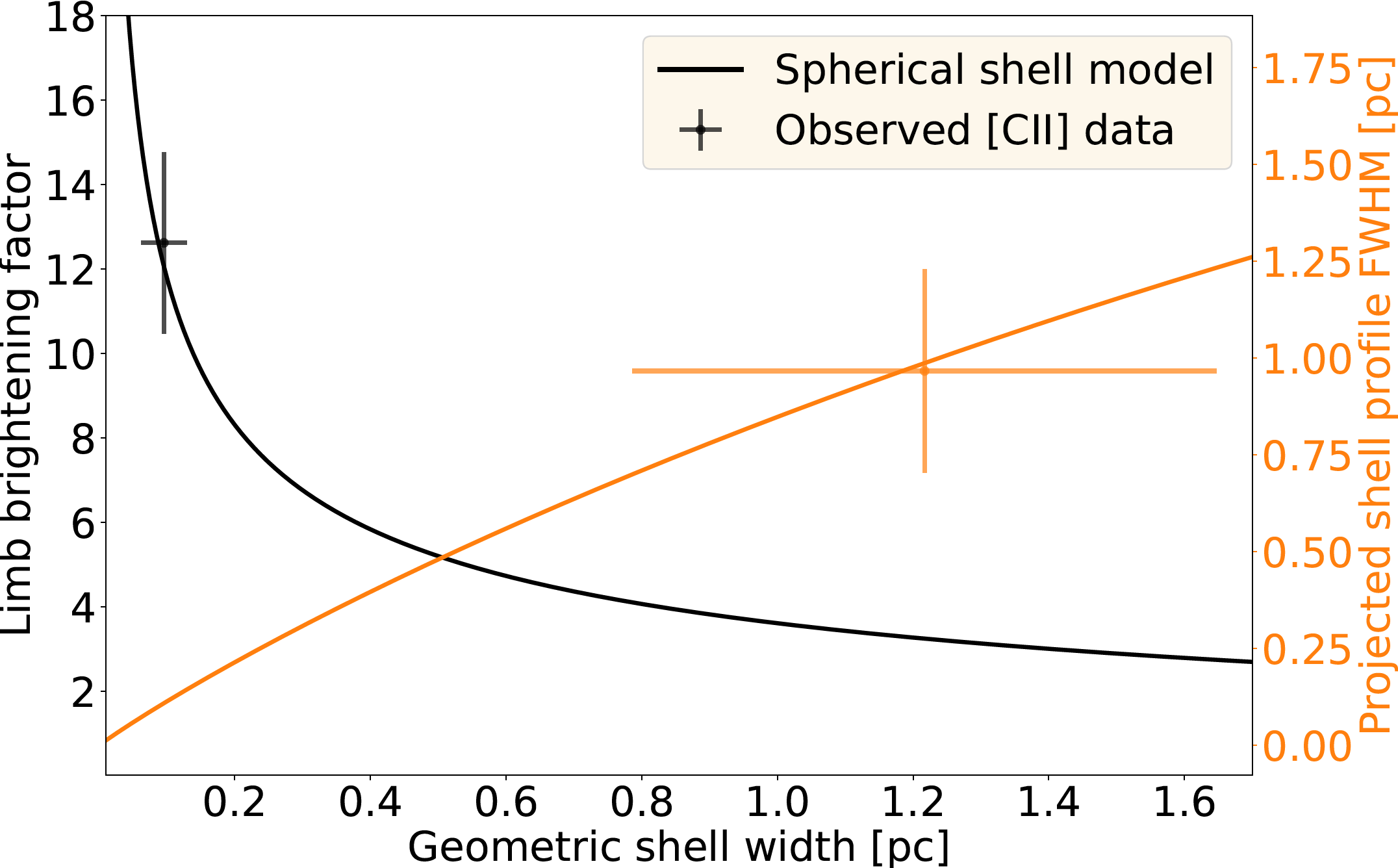}
    \caption{Limb brightening factor and projected shell FWHM assuming a bubble geometry with a radius of 6.5 pc for the total intensity of RCW~79, based on the formalism in \citet{Kabanovic2022}. To be a self-consistent spherical bubble, the deduced geometric shell width from the limb brightening and the projected shell FWHM must be the same within the error, which is not the case. This shows that the emission distribution of RCW~79 is not consistent with a spherical shell morphology. In this plot the results for the highest projected shell FWHM (1.8 pc and 2.3 pc) were not included as the FWHM formalism does not allow a solution for these values.}
    \label{fig:limbFWHM}
\end{figure}

The sketch in Fig. \ref{fig:limb} illustrates the limb brightening effect in a spherical shell. For the calculations we assume that the observed line is optically thin such that the observed intensity is given by S$_{\nu}\tau_{\nu}$ = S$_{\nu}\kappa_{\nu}\Delta$y = j$_{\nu}\Delta$y, where S$_{\nu}$ is the specific intensity, $\tau_{\nu}$ the optical depth, $\kappa_{\nu}$ the attenuation coefficient, and $\Delta$y the distance. 
At the center  of the shell (x=0) there is no limb brightening, which results in an observed intensity of j$_{\nu}$r. At the position with peak limb brightening, the observed intensity is j$_{\nu}$y$_{peak}$. At this position, y$_{peak}$ can be described as
\begin{equation}
    y_{peak} = R\sqrt{1 - \left( \frac{R-r}{R} \right)^{2}}
\end{equation}
using the Pythagorean theorem,  where R is the radius of the shell and r the 3D thickness of the shell.
The observed limb brightening is thus given by
\begin{equation}
    \frac{I_{peak}}{I_{cent}} = \frac{R}{r}\sqrt{1 - \left( \frac{R-r}{R} \right)^{2}}
.\end{equation}
In an isolated region, the observed shell width would be simply given by r as there is no emission outside the shell. 
However, shells are embedded in interstellar gas which means that the above definition is not always applicable. 
Therefore, we decided to define the width at half of the maximum intensity which is defined by $\Delta$x$_{WHM}$ in Fig. \ref{fig:limb}. We do not employ the full width at half     maximum (FWHM) because the projected shell intensity profile is not symmetric around the peak emission (see Fig. 14 in \citealt{Kabanovic2022}). The associated length through the shell is then given by y$_{WHM}$ = y$_{peak}$/2 so that $\theta_{WHM}$ is defined by
\begin{equation}
    \theta_{WHM}= {\rm sin}^{-1}\left( \frac{1}{2}\sqrt{1 - \left( \frac{R-r}{R} \right)^{2}} \right)
.\end{equation}
On the other hand, we also have
\begin{equation}
    R \, {\rm cos}\theta_{WHM} = R - r + \Delta x_{WHM}
.\end{equation}
Using the relation
\begin{equation}
    {\rm cos(sin}^{-1}(\phi)) = \sqrt{1-\phi^{2}}
,\end{equation}
we then obtain
\begin{equation}
    R - r + \Delta x_{WHM} = R\sqrt{\frac{3}{4} + \frac{1}{4}\left( \frac{R-r}{R} \right)^{2}}
,\end{equation}
which results in
\begin{equation}
    \frac{\Delta x_{WHM}}{R} = \frac{r}{R} - 1 + \sqrt{\frac{3}{4} + \frac{1}{4}\left( \frac{R-r}{R} \right)^{2}}
.\end{equation}
These equations are also employed in the analysis of \citet{Bonne2023a}. The observed I$_{peak}$/I$_{cent}$ and $\Delta$x/R thus provide two independent methods for calculating r, using Brent's method,\footnote{https://docs.scipy.org/doc/scipy/reference/generated/scipy.optimize.brentq.html} and for evaluating whether the results are consistent (which is expected for a spherical expanding shell). To   this end, we created velocity-integrated intensity cuts over the center of the RCW~79 \HII\ region for the blue- and redshifted wing and the total velocity range (-25~km~s$^{-1}$ to -70~km~s$^{-1}$); this was done in intervals of 45$^{o}$. 
To make sure we were not working with upper limits, we only worked with the cases where I$_{cent}$ $\ge$ 3$\sigma$. To determine $\Delta$x$_{WHM}$ we focused on the pixels with a higher radius than the peak intensity to find the first pixel with an observed intensity $<$I$_{peak}$/2. With linear interpolation between this and the previous pixel, we determined $\Delta$x$_{WHM}$. 
Based on these observed values we then inferred the actual shell width (r). The results are summarized in Table~\ref{tab:shellWidth}; they show that the resulting shell width (r) from the limb brightening and $\Delta$x$_{WHM}$ measurements are inconsistent. To explain the observed limb brightening, a significantly smaller width of the shell would be required than the observed width of the shell. We also note that this should not be a spatial resolution effect since our spatial resolution of $\sim$0.4 pc is smaller than the observed half width of the shell ($\Delta$x$_{WHM}$).\\ 
The model we presented here is based on three assumptions: the emission is optically thin, the shell has a constant local emissivity (j$_{\nu}$), and the shell is spherical. Optical depth effects for the limb brightening would give results opposite to the observations. Therefore, the high-velocity emission has to be highly inhomogeneous, and thus is likely non-spherical. Based on the very bright \CII\ emission at the edge of the \HII\ region, an elegant solution would be that the high-velocity gas is predominantly gas ablated from the edge of the ring. This appears to fit with the observed kinematics, and would also provide a solution for the timescale issue.\\ 
Recently, the same model was also discussed in \citet{Kabanovic2022} in the context of RCW~120. We independently confirmed our result for RCW~79 with the formalism for the geometry of a spherically expanding bubble presented in \citet{Kabanovic2022}. Figure \ref{fig:limbFWHM} shows the results from this independent analysis and demonstrates again that the observed limb brightening in RCW~79 is inconsistent with the observed shell width. Although in essence the model of \citet{Kabanovic2022} is the same as the one above, it builds on the observational definition of a FWHM (instead of WHM in the formalism above) for the observed shell width. Applying this formalism thus verifies that the results above are not the result of a bias in our observational definition.

\begin{table}[]
    \centering
    \caption{Inferred shell widths for RCW~79}
    \begin{tabular}{ccc}
    \hline
    \hline
         & r$_{I_{peak}/I_{cent}}$ & r$_{\Delta x_{WHM}}$ \\
         \hline
        Total & 0.05-0.13 pc & 0.5-1.1 pc\\
        Blue wing & 0.08-0.14 pc & 0.7-0.8 pc\\
        Red wing & 0.26-0.93 pc & 0.7-1.1 pc\\
        \hline
    \end{tabular}\\
    {{\bf Note:} The inferred width of the shell (r) in \CII\ based on the limb brightening (r$_{I_{peak}/I_{cent}}$) and the observed half width of the shell (r$_{\Delta x_{WHM}}$). The results based on the two methods are clearly inconsistent.}
    \label{tab:shellWidth}
\end{table}

\section{Column density in the \CII\ high-velocity wings}\label{sec:appB}
The C$^{+}$ column density can be estimated in each velocity channel using the equation from \citet{Goldsmith2012}
\begin{equation}
    \Delta T_{A} = 3.43\times10^{-16} \times\left[ 1 + 0.5\times{\rm e}^{91.25/{\rm T_{kin}}} \left( 1 + \frac{2.4\times10^{-6}}{{\rm C_{ul}}} \right) \right]^{-1} \frac{{\rm N(C^{+})}}{\delta {\rm v}}
.\end{equation}
Here T$_{kin}$ is the kinetic temperature, N(C$^{+}$) the C$^{+}$ column density, $\delta$v the spectral bin width, and C$_{ul}$ the collisional de-excitation rate given by 
\begin{equation}
    C_{ul} = n \times R_{ul}
,\end{equation}
where n is the density and R$_{ul}$ is the de-excitation rate coefficient for atomic hydrogen given by
\begin{equation}
    R_{ul} = 7.6 \times 10^{-10}\,cm^{3}\,s^{-1}\,(\frac{T_{kin}}{100\,{\rm K}})^{0.14}
.\end{equation}
Based on the results of the PDR Toolbox for the FUV field strength, we use T$_{kin}$ values between 50 and 250 K. The quoted mass, momentum, and energy ejection intervals in the paper are based on this temperature range. As there is no direct estimate of the density, we assume n$_{H}$ = 10$^{3}$ cm$^{-3}$. This value is reasonable seeing $^{12}$CO is basically undetected in the high-velocity wings (but does show some hints of localized clumps), and it is close to the predicted density for the PDR assuming thermal equilibrium with the ionized gas phase. Using lower density values would only lead to higher mass, momentum, and energy ejection rates. To convert the C$^{+}$ column density to hydrogen column density we used [C$^{+}$]/[H] = 1.6$\times$10$^{-4}$ from \citet{Sofia2004} and a mean atomic mass of 2.2$\times$10$^{-27}$~kg. Based on the column density maps, we can calculate the total mass associated with the blue- and redshifted high-velocity wings. For temperatures between 50 and 250 K, this results in masses between 5.9$\times$10$^{3}$ M$_{\odot}$ and 1.5$\times$10$^{3}$ M$_{\odot}$, respectively, for the blueshifted gas and 6.5$\times$10$^{3}$ M$_{\odot}$ and 1.6$\times$10$^{3}$ M$_{\odot}$, respectively, for the redshifted gas.\\
To estimate a timescale,  which is necessary to calculate the mass, momentum, and energy ejection rates,  we work with the observed LOS velocity relative to the bulk motion of the cloud and the estimated typical distance in the LOS (i.e., 5 pc). Lastly, it should be noted that the calculated mass, momentum, and energy ejection rates have several factors of uncertainty, for example  projection uncertainties, density uncertainty, and  the assumed [C$^{+}$]/[H] value. This makes it difficult to estimate the error bars, but suggests that   at least   a factor two should be considered for the uncertainty. As the calculations are plagued by many uncertainties, future work will explore numerical simulations to estimate correction factors for potential biases.\\
Lastly, we note that we verified whether the high-velocity \CII\ emission could originate from the \HII\ region instead of the PDR. To do this, we ran several constant pressure CLOUDY models \citep{Ferland2017} for a \HII\ region resembling RCW~79, using the O-star luminosity from \citet{Martins2010}, a blackbody radiation spectrum of 3.9$\times$10$^{4}$ K, radii for the \HII\ gas between 1 and 5 pc, and Orion nebula abundances. This showed that $<$14\% of the total \CII\ emission originates from the \HII\ region itself. This implies that $<$30\% of the emission in the high-velocity \CII\ wings can be attributed to the \HII\ region. \CII\ excitation in the ionized gas phase can give rise to slightly lower masses in the high-velocity wings per unit of emission (by a factor 2-3). However, as the \CII\ emission coming from the ionized gas phase is limited, this would lead to a correction of  $<$15\%, which is well below the  other uncertainties that go into the calculation with currently available data.


\end{appendix}

%
%


\end{document}